\documentclass{article}

\usepackage{arxiv}
\usepackage[utf8]{inputenc} % allow utf-8 input
\usepackage[T1]{fontenc}    % use 8-bit T1 fonts
\usepackage{url}            % simple URL typesetting
\usepackage{booktabs}       % professional-quality tables
\usepackage{amsfonts}       % blackboard math symbols
\usepackage{nicefrac}       % compact symbols for 1/2, etc.
\usepackage{microtype}      % microtypography
\usepackage{graphicx}
\usepackage{doi}
\usepackage{soul} 
\usepackage{color}
\usepackage{enumerate}
\usepackage{hyperref}
\usepackage{natbib}
\setcitestyle{numbers}

\usepackage{xcolor}

\hypersetup{
pdftitle={Knowledge mining of unstructured information: application to cyber-domain},
pdfsubject={cs.CL, cs.IR, cs.LG},
pdfauthor={Tuomas Takko, Kunal Bhattacharya, Martti Lehto, Pertti Jalasvirta, Aapo Cederberg, Kimmo Kaski},
pdfkeywords={knowledge graphs, knowledge mining, cyber intelligence},
}

\title{Knowledge mining of unstructured information: application to cyber-domain}

\author{Tuomas Takko \\
Department of Computer Science\\
Aalto University School of Science\\
00076, Finland \\
\texttt{tuomas.takko@aalto.fi}
\And
Kunal Bhattacharya \\
Department of Industrial Engineering and Management\\
Department of Computer Science\\
Aalto University School of Science\\
00076, Finland \\
\And
Martti Lehto \\
Faculty of Information Technology\\  
University of Jyväskylä\\ 
PO Box 35, 40014, Finland\\
\And
Pertti Jalasvirta \\
Cyberwatch Finland Oy\\ 
Tietokuja 2, 00330, Finland\\
\And
Aapo Cederberg \\
Cyberwatch Finland Oy\\ 
Tietokuja 2, 00330, Finland\\
\And
Kimmo Kaski \\
Department of Computer Science\\
Aalto University School of Science\\
00076, Finland \\
The Alan Turing Institute\\
96 Euston Rd, Kings Cross, London NW1 2DB, UK
}

\begin{document}
\maketitle

\begin{abstract}
Information on cyber-related crimes, incidents, and conflicts is abundantly available in numerous open online sources. However, processing the large volumes and streams of data is a challenging task for the analysts and experts, and entails the need for newer methods and techniques. In this article we present and implement a novel knowledge graph and knowledge mining framework for extracting the relevant information from free-form text about incidents in the cyberdomain.
The framework includes a machine learning based pipeline for generating graphs of organizations, countries, industries, products and attackers with a non-technical cyber-ontology.
The extracted knowledge graph is utilized to estimate the incidence of cyberattacks on a given graph configuration.
We use publicly available collections of real cyber-incident reports to test the efficacy of our methods. The knowledge extraction is found to be sufficiently accurate, and the graph-based threat estimation demonstrates a level of correlation with the actual records of attacks.
In practical use, an analyst utilizing the presented framework can infer additional information from the current cyber-landscape in terms of risk to various entities and propagation of the risk heuristic between industries and countries. 

\end{abstract}

%.
\thispagestyle{empty}

\noindent
\section{Introduction}

The cyberspace is increasingly facing challenges in the form of  persistent and devious threats from state and non-state actors alike. Given the growth of smart devices, data storage options, and supply-chain dependencies, establishing the security and resiliency in the cyberdomain has become an imperative for companies and organizations across sectors~\cite{weforum2021}. A key challenge here is to assimilate the large-scale data in free-forms, such as reports on incidents and vulnerabilities, that is openly available from numerous sources including vulnerability databases and international agencies~\cite{certEU}. An ad-hoc structuring of information by interlinking reports on events, alternately a knowledge graph framework~\cite{liu2022recent, li2019knowledge,piplai2020creating,li2020cskb}, appears to be a viable solution. The concept of knowledge graphs has been adopted for structuring and processing of the technical information on known vulnerabilities, malicious IP addresses and different relevant threats in the cyberdomain, as well as for associating other related entities such as software developers.

% Move to related work if possible
In this article we present a strategic level framework for analyzing the cyber-landscape by utilizing publicly available textual reports of various incidents. The objective being a broad yet condensed view of the relevant interconnected entities and the prevailing threats. In addition to a visual tool, this article explores the usability of the resulting knowledge graph in terms of estimating the risk of cyberattacks.
Such frameworks have been proposed in the past, but only a few studies provide methods for practical and automated construction of visual and graph-based solutions. The framework is based on high level descriptions of cyber-incidents, like attacks and breaches, which are easily obtainable from open online sources. %

This work shares a broader objective similar to the work of \citet{bohm2018graph}, in which the authors described and justified a human-readable and visual approach for analyzing complex cyberattack reports. As the success of security experts depends on the readability of available intelligence, information in structured formats and ontologies, require additional tools and frameworks for actionable usage. In general, ontologies such as STIX~\cite{barnum2012standardizing}, UCO~\cite{syed2016uco}, and STUCCO~\cite{iannacone2015developing} consist of various technical or higher level entities and their possible interrelations, that facilitates the interlinking of entities and events. While these ontologies have excelled in focusing on the microscopics and different technical sophistications, there still remains ample scope to portray the cyber-landscape in a clear and readable manner for the ease of analysis and subsequent decision making at a strategic level.

In the vein of earlier research~\cite{joshi2013extracting, li2020cskb, bohm2018graph}, we extend the knowledge graph constructed from unstructured data by joining information from separate other sources of data. We use crawling and querying for additional records and information about the entities from sources such as DBpedia \cite{10.5555/1785162.1785216}. The additional information is aimed to sufficiently populate the ontology and to introduce interconnectedness in the knowledge graph which later allows us to calculate the risk.
We demonstrate that the graph can be used to determine a risk level for the entities in the graph by using historical data on cyberattacks. This risk level could be used to estimate the likelihood of future cyberattacks given the past incidents on connected entities. 

This article is structured in the following manner. In Section 2, we establish the position of the current framework in relation to existing works and studies in the field of open source and knowledge graph based systems, focusing on the studies that have overlapping methods or data sources.
Next, in Methods and Materials (Section 3) we describe the processing pipeline for producing a strategic level knowledge graph from unspecified textual sources, such as news reports on cyberattacks.  In Section 4, we analyze the knowledge graph and use it to measure a type of risk. Finally, in Section 5 we discuss the relevance of our findings in terms of their usefulness to cyber-analysts and enumerate the limitations. In Section 6 we summarize our findings and discuss the possible future improvements.

\section{Related Work}

The concept of knowledge graph, where complex information is represented as nodes and edges with semantic relations \cite{ehrlinger2016towards,duan2017specifying}, has become increasingly popular in numerous fields of research and in the implementation of information-driven applications.
Improved methods for extracting meaningful information and entities from unstructured text, see e.g. \cite{finkel2005incorporating}, as well as the increasing coverage of linked data from various endpoints (such as DBpedia \cite{10.5555/1785162.1785216}) has made it possible to query for relevant information and to connect information from text to existing records of various entities, events and items.
The applications of knowledge graph ranges from systems in healthcare \cite{shen2017constructing, rotmensch2017learning} to search systems and scientific document indexing \cite{auer2018towards}.

In cybersecurity and cyber intelligence, the use of knowledge graphs and linked data has been prevalent due to the mostly structured nature of the recorded data related to intrusion detection systems, software vulnerabilities and malicious actors~\cite{georgescu2017using, liu2022recent}.
For instance, online databases like NVD \cite{nvd}, CVE\cite{cve} and CWE\cite{cwe} provide regular updates on software and system vulnerabilities on a structured format.
Cyberdefense benefits from synergy and cooperation, but sharing and interpreting various threat intelligence reports and databases requires standardized formats and protocols for the analysts to have a common language \cite{mavroeidis2017cyber}.
Thus, there has been extensive research done for constructing taxonomies and ontologies to standardize the formats of linked data on threat intelligence such as software and system vulnerabilities, malware \cite{rastogi2020malont}, and attacks in general \cite{syed2016uco, barnum2012standardizing}.
Using these types of ontologies to provide formalism and structure, various framework-type approaches to situational cyber awareness have been developed, for instance for different vulnerabilities, assets and network topologies during cyberattacks \cite{li2020cskb, komarkova2018crusoe, heinbockel2016mission, noel2016cygraph, bohm2018graph}.  
Other approaches for extracting relevant information on cyberattacks and vulnerabilities from different unstructured text sources, such as social media, have been used as early warning signals for cyber-risks \cite{schafer2019blackwidow, tavabi2018darkembed,mittal2016cybertwitter, mittal2016cybertwitter, mittal2019cyber, neil2018mining, jia2018practical}.

This study is related to earlier works~\cite{joshi2013extracting,li2020cskb,kejriwal2017information} that describe and implement methods for a pipeline with the objective to turn unstructured data into knowledge graphs based on specific ontologies. For instance, \citet{joshi2013extracting} described a framework that processes unstructured web text from security bulletins and blogs alongside with the vulnerability data from the NVD, CVE and CWE datasets, recognizes the entities and concepts, and finally connects them by using information from DBpedia Spotlight. 
%This data is then processed into triples for constructing a knowledge graph that enables automatic consumption of the threat landscape.
In another work, \citet{li2019knowledge} proposed a method for building a  knowledge base with similar rules and structure. But instead of considering the software and hardware vulnerabilities, the ontology used device properties, attack properties and attack features. 
The datapoints were gathered from the network level information using a neural network classifier. A third study that is relevant here is by \citet{kejriwal2017information} that described an information extraction method for unstructured text, scraped from illicit web domains. The authors proposed methods for annotating and extracting information such as entities and locations using unsupervised methods based on an initially annotated corpora.

The framework presented in this paper shares principle level similarities to the studies described above, in terms of the structure of data processing pipeline and the methods.
%The general objective is also similar to ours, i.e. to process unstructured text scraped from various unknown format news sources into a knowledge graph.
This shares similarities to the general objective of the work by \citet{li2020cskb} in portraying cyberattacks in a knowledge graph format.
The framework in the paper by \citet{li2020cskb} processes data from the network and information systems of an entity, whereas the strategic domain approach of this study is restricted to open source information from security bulletins and news sources, thus limiting the number of features and amount of information available. 
The information extraction part of this study has principles similar to the work of \citet{kejriwal2017information}, with the objective of processing information from unknown domains and extracting the relevant entities and their relationships. 
We extract the entities using a named entity recognizer (NER) from spaCy~\cite{spacy} and compare the extracted relevant entities to the knowledge base of DBpedia using DBpedia Spotlight, similar to \citet{joshi2013extracting}.
While sharing similarities and overlaps to previously described work, the framework in this paper is dependent only on the most surface level descriptions of the events, which can be acquired from easily available sources online.
The level of observing and mapping the various cyberattacks takes place on a non-technical and strategic level, while the technical level can in the future be connected to these events if the respective information is available.

\section{Methods and Materials}

\begin{figure}
    \centering
    \includegraphics[width=0.9\textwidth]{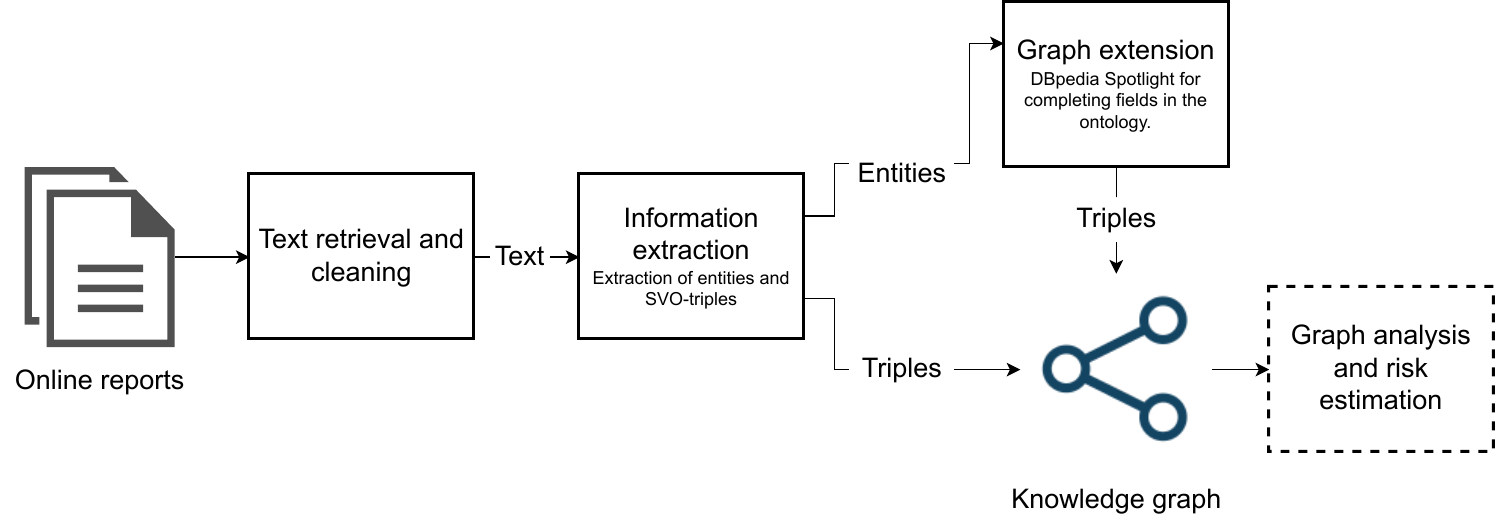}
    \caption{Process pipeline of the proposed knowledge mining framework. The framework and the modules are depicted as boxes with the correct order. The pipeline retrieves, cleans and extracts information from unstructured text and computes graph-level features for the analyst to investigate in the final knowledge graph.}
    \label{fig:pipeline}
\end{figure}

Framework for processing unstructured information consists of three distinct modules, namely an information retrieval module, an information extraction module and finally a module for risk measurement and graph analysis. 
Even though the source material consists of written news and reports, the framework implementation does not reuse the text or otherwise infringe the copyright of the authors of the original pieces. 
The process with the corresponding modules is shown in Fig. \ref{fig:pipeline}.
The first module aims to gather and process relevant unstructured information from unspecified online sources. 
It begins by collecting a list of urls of news reports of cyberattacks that are of interest to the analyst.
The module proceeds by requesting the page from the given url, if the source allows software agents, and cleaning the obtained text by removing irrelevant content such as html-tags, other urls or embedded content.
This cleaned text is then processed by removing stop words and extracting the relevant entities and their relationships in the information extraction module.
The relationships between the target and the attacking entities are extracted as a triple in the form of ``target - attackedBy - attacker''.
The extracted entities are compared to the results of DBpedia Spotlight \cite{isem2013daiber}, which finds related records in DBpedia \cite{10.5555/1785162.1785216} as linked data, which we then use to complete the fields in the ontology.
DBpedia Spotlight annotates the entities found in the text and performs disambiguation using the context of the phrases.
In an ideal situation, these entities are correctly resolved and found in DBpedia, but in a situation where this additional information is not found, we omit the information while keeping the entity as it was recognized by the spaCy NER \cite{spacy} and adding the triple of attacker-victim relationship.
In a complete system one could also crawl other sources for additional information, such as software vulnerabilities.
Lastly, we use the generated knowledge graph for estimating a naive measure for risk.
The risk level in this study is based on the frequency of attacks in the connected entities of the knowledge graph.

%Practical implementation in short
These modules were implemented in Python 3.7 using libraries for web scraping, spaCy \cite{spacy} for information extraction and NetworkX \cite{hagberg2008exploring} for graph analysis.
For the purpose of demonstrating our approach in this study, we opted to use the openly available datasets of cyberattacks from Hackmageddon \cite{Hackmageddon} containing reported attacks from year 2017 to 2020.
The human-annotated dataset contains descriptions of targets, attackers, attack types, dates, countries and links to the original reports, which we use to obtain the full text report. 
The remaining fields are used in the evaluation of information extraction methods of this framework as well as to substitute for missing relationships from DBpedia.
In a real use-case an analyst would use their own news sources or knowledge bases and use the framework via a user interface, or other applicable method.
However, for the sake of clarity we restrict the number of cases analyzed in the knowledge graph to the contents of the Hackmageddon dataset.
Some organizations might experience attacks or preparations of an attack on a daily basis, but those are not reported in the news whether due to their commonality, minor damage or because the organization is not releasing the information.
The attacks recorded in this set of data are the ones where the attack itself is already operational and deemed news-worthy.

\subsection{Information Extraction} 

The primary aim of this module is to identify the victim and the perpetrator of an attack for a given piece of text. The approach followed here generally belongs to relation extraction methods used for constructing cybersecurity knowledge graphs~\citet{pingle2019relext}, however, this is an unsupervised method~\cite{rosenfeld2006ures}. The approach comprises of extraction of subject-verb-object (SVO) triples, scoring for named entities, and ranking of entities. The SVO triples are extracted using a mixture of rule based methods~\cite{michael2019ICDM} and parsing of the dependency tree~\cite{d2018parser}. Below we will broadly describe the steps different steps in the approach. The finer details of the method and related concepts will be reported elsewhere.

\begin{enumerate}[(i)]
\item First, the subjects and objects are tagged using spaCy noun phrases (noun-chunks).
\item The verb phrases are identified as the most general pattern: particle + adposition + verb/auxillary + particle + adposition + adjective/adverb + adposition. Similarly, lone adpositions, adverbs and adjectives are tagged. To take into account complex predicates, light verb constructions that include nouns within are also recognized, for example, the phrase `gained access to'~\cite{etzioni2011open}. Additionally, Hearst patterns~\cite{hearst1992automatic} are identified from a pre-compiled list and using the patterns nouns are linked by using the dependency tree structure parsed by spaCy.
\item A coarser dependency tree is constructed using the noun and the verb phrases and the original dependency structure. Note that the dependency parsing may vary depending on the language model used by spaCy. Using this coarser tree and considering the subject-verb-object order we generate the triples. The tree is parsed such that the conjugated verbs are crawled and associated with all subjects and objects. 
\item A co-reference resolution for the set of noun phrases is performed using the package NeuralCoref~\cite{wolf2017state}.The resulting output from the resolution are clusters of noun phrases, where each cluster implies a single co-referenced mention. 
\item Named entities are tagged using the spaCy named entity recognizer.
\item A map between the named entities in the text and the clusters from co-reference resolution is created. Using the map the subjects and the objects in the triples are replaced with the named entities. 
\item For each named entity a `target score' score is calculated in the following fashion. A list of attack tokens is considered. The list is populated with a set of seed tokens, such that `hacked', `breached', etc., and further are extended by including the inflections. Given an SVO triple we check for the presence of an attack token inside the verb phrase. If a token is found and the triple has an active voice then the entity corresponding to the object gets its target score incremented by $+1$. If the voice is passive, the target score corresponding to the entity in the subject is incremented. The process is repeated for all the triples and the final target scores are obtained.
\item The number of occurrences of each entity and the order in which they appear in the text are also calculated. For each entity, a compound score is calculated by adding the min-max normalized values of the target score, the frequency of appearance, and the order (reversed). The entities are ranked in descending order of the compound score. The topmost identity is identified as the primary target mentioned in the text. 
\end{enumerate}

\begin{figure}
    \centering
    \includegraphics[width=0.9\textwidth]{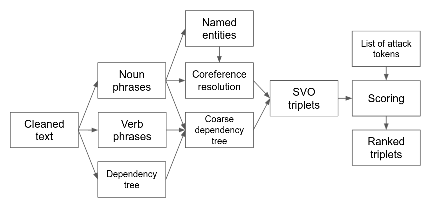}
    \caption{The components of the process of information extraction are shown. A major part of the process is implemented using the Python spaCy library.}
    \label{fig:nlp-flow}
\end{figure}

The above method is illustrated in Fig.~\ref{fig:nlp-flow}. We find that the above method yields an accuracy of 60$\%$ for the top-most ranked entity to be the true target. However, the accuracies for the true target to be in top-2 and top-3 ranked entities are 75$\%$ and 83$\%$, respectively.  
If solely the frequency or the order of appearance is taken into account the accuracy for the true target to be in the top-most and top-3 entities are around 50$\%$ and 70$\%$, respectively. 
Note that in general a news piece has 10-20 entities, and therefore, a baseline accuracy would be much lower in comparison. For determining the attacker, we follow a scoring method similar to (vii) by choosing the subject (object) in a triple when the voice is active (passive). However, for most of the instances in the dataset the identity of the attacker is not known. Therefore, in this case the accuracy can not be ascertained.  
In our future work we will provide methods whereby models can be trained on linguistic features, and quantities like frequency and order.

\subsection{Domain Ontology Structure}
\begin{figure}[ht]
\begin{minipage}[c]{0.55\linewidth}
\centering
\includegraphics[width=1\textwidth]{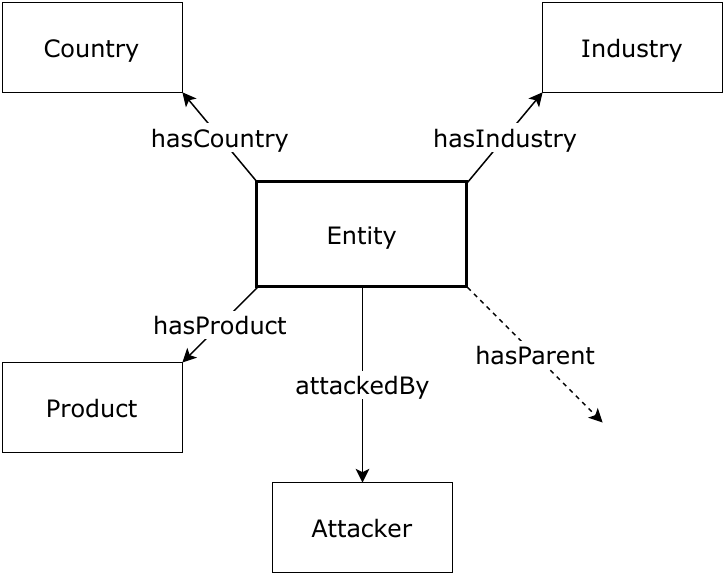}
\end{minipage}
\begin{minipage}[c]{0.4\linewidth}
\centering
\resizebox{\linewidth}{!}{%
\begin{tabular}{|c|c|}
    \hline
    \textbf{DBpedia}  & \textbf{Study}  \\ \hline
    \hline
    foaf:name & Name \\ \hline
    dbo:industry & hasIndustry  \\ \hline
    dbp:locationCountry & hasCountry  \\ \hline
    dbo:product & hasProduct \\ \hline
    dbo:parentCompany & hasParent \\ \hline
\end{tabular}}
\end{minipage}\hfill

\caption{A novel strategic level cyber ontology. Subjects and objects are entities (boxes) that are connected via their relative predicates (arrows). The labels tell the class names in the ontology.
    The relationships between the extranous data from DBpedia (labeled DBpedia) and the strategic-level ontology presented in this study (labeled Study). The triple describing a cyberattack is present only in the novel ontology of this study. The hasParent relationship is from an entity to another entity.
    }\label{fig:onto}

\end{figure}

The framework of this study utilizes a novel domain ontology for defining the elements and the relationships appearing in the knowledge graph.
The ontology is depicted in Fig. \ref{fig:onto} and it captures knowledge on the entities and actors at a strategic level, i.e. at a level that describes real world structures and helps in constructing a broader picture of the whole field at once.
The extracted information for each report on a cyberattack depicts the main attributes of an organization and ideally forms a connected network, in which visualizing trends and campaigns along with individual attack incidents is possible.
The entities, such as companies and organizations, are described by their countries and industries as well as by their products and possible child-parent relationships to other entities.
Different countries, products and industries appear in the knowledge graph as nodes alongside the organizations and attacking entities.
We categorize industry and country nodes into central nodes and rest of the nodes non-central. 

As we are using DBpedia Spotlight to obtain information on the extracted entities, the ontology can be considered to share predicates with the ontology of DBpedia.
The relationships and their counterparts in DBpedia's syntax are depicted in the table in Fig. \ref{fig:onto}.

When populating the knowledge graph using this ontology, we are not setting any requirements or rules to the types of categories that might appear in the automated construction process.
Every entity is considered as a type of organization, with the distinguishing feature being the type of industry the entity has.
For instance, a government organization would have an industry indicating public service.
The set of entity attributes for describing cyberattacks and events in our ontology were chosen as such to maintain readability and simplicity of the knowledge graph.
It is also worth noting, that increasing the number of predicates for a given entity also affects the network properties of the resulting knowledge graph.
Naturally, the number of these predicates can be increased if the analyst requires other information within the boundaries of information available, but the current set serves as a backbone for the purpose of this study.
The finalized result of the knowledge graph using the ontology presented here contains five types of nodes (entity, country, industry, product, attacker) and the five relationships described above.
An example of a subset of the resulting knowledge graph is shown in Fig. \ref{fig:case}.

\subsection{Measuring Risk Level} 
In addition to the situational awareness and human readability provided by the knowledge graph, we aim to quantify risk for the entities in the graph, but rather than trying to predict the occurrence of cyberattacks, the focus is on measuring risk from the relationship between sequential attack records.
The level of risk in this study is measured from the network structure of the knowledge graph.
As an attack is recorded as the SVO-triple and the recorded date, we can construct risk levels for the central nodes, which consist of industry nodes and country nodes.
The risk levels for the central nodes in the knowledge graphs are then used as a proxy for the connected entities via the linkages in the network structure. 

In this initial model, we consider the risk $r$ for a single central node $c$ to be given by a sum of decaying exponential weights over the past events, such that
\begin{equation}
    r_c(t) = \sum_{i}^{t} e^{i-t},
\label{Eq:centralrisk}
\end{equation}
where $t$ denotes the time when the risk is calculated, and $i$ is a time when an attack happened. The unit of time is considered as a parameter for adjusting the contributions of the past events to the current risk. After testing on our data for different values, we chose time to be in units of 30 days.  
For each attack-triple the time of reported occurrence is stored in the data structure.%, but not visualized in the knowledge graph.
Also, should an attack be spread on multiple days, each day of the attack would be presented as a separate triple.

The time step can be chosen for an appropriate duration, considering the type of data represented in the knowledge graph.

We also calculate the second central neighbors for the entity nodes by constructing a projection (see Fig. \ref{fig:hacknet}) of the network in such a way that the central nodes sharing entity nodes are connected in the projection.
In addition to forming connections, the projection can be used to provide weights on the links between the central entities based on the number of shared entities. 
In the scope of this study we chose to consider every link with equal weight due to the fact that the projection should be temporal and change over time in terms of the evolving amount of common entities between the central nodes.
Constructing the projection allows us to investigate whether the risk propagates across the network and whether certain types of nodes have more importance when considering the weights in the risk measures.

For a non-central entity $e$ in the graph (i.e. an organization or a company), the risk level at a certain time step can be calculated from the neighboring central nodes by calculating a sum of the means
\begin{equation}
    r_e = \overline{r_e(C)} + \overline{r_e(I)} + \overline{r_e(c)} + \overline{r_e(i)},
\label{Eq:risk-variables}
\end{equation}
where  $\overline{r_e(C)}$ denotes the mean of risk for the first neighbor country type nodes, $\overline{r_e(I)}$ denotes the mean of risk for first neighbor industry nodes and $c$ and $i$ denote the risk for the second neighbor countries and industries in the projection, respectively.
The second neighbors in this measure are considered to be the immediate neighbors of the central nodes $C$ and $I$ in the projection, $C$ and $I$ being connected to the focal entity $e$ in the knowledge graph.
We construct these risk measures into a dataset, in which for each day any non-central entity can be evaluated using a vector of these four values.

\begin{figure}[ht]
    \centering
    \includegraphics[width=0.8\textwidth]{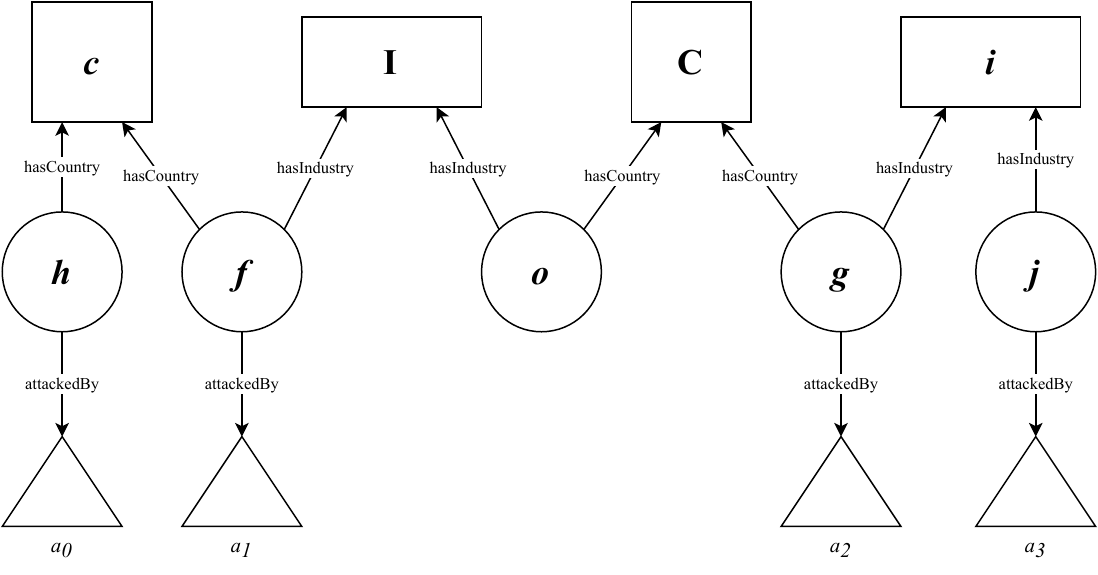}
    \caption{Example of the structure of the knowledge graph for calculating the heuristic risk value for the focal entity $o$. The focal entity is connected to industry $I$ and country $C$. The rest of the graph is populated with entities $h$, $f$, $g$ and $j$ as well as industry $i$ and country $c$. The recorded attacks to the entities are denoted from $a_0$ to $a_3$ in the order of occurrence at times $t_0$ to $t_3$.}
    \label{fig:exampleKG}
\end{figure}

% call the risk a heuristic?
In practice, the risk level shown in Eq.~(\ref{Eq:risk-variables}) is constructed as follows.
Let us consider an organization $o$ that is connected to a single industry node $I$ and a single country node $C$ in a random knowledge graph at time $t$. 
The structure of this example graph is shown in Fig. \ref{fig:exampleKG}. 
To calculate the sum for $r_o$ we first consider the mean of the risk value depicted in Eq.~(\ref{Eq:centralrisk}) for the first neighbour nodes $I$ and $C$.
As the attacks to entities linked to the nodes $I$ and $C$ occur at times $t_1$ for attack $a_1$ and at $t_2$ for attack $a_2$, the risk values are $r_o(C) = e^{t_2 - t}$ and $r_o(I) = e^{t_1 - t}$, respectively.
In a similar fashion, we can calculate the risk for the second neighbour nodes, i.e. the nodes that share entities with the central nodes the focal entity is connected to, $i$ and $c$.
The risk measures for these nodes would be $r_o(c) = (e^{t_0-t}+e^{t_1-t})/2$ and $r_o(i) = (e^{t_2-t}+e^{t_3-t})/2$.
Thus, the risk for entity $o$ would be $r_o = e^{t_2 - t} + e^{t_1 - t} + (e^{t_0-t}+e^{t_1-t})/2 + (e^{t_2-t}+e^{t_3-t})/2$.
This value of the risk not a measure of probability.
We investigate the usefulness of this measure by using the components of the sum as a set of values.

%% add figure for the example entity etc

\subsection{Experimental setup and analysis methods}
% Explain and establish the methods used in the evaluation and investigation of the dataset KG
We evaluate the knowledge mining framework by combining a dataset from Hackmageddon's cyberattack timelines for each month from January 2017 to April 2021.
For each article we crawled the original text whenever possible and processed its text through the NLP pipeline, extracting the attacked entity, the attacking entity and matching them with the entities recognized by DBpedia Spotlight. 
As the empirical data contains these fields already annotated by humans, we also compared the extracted entities to the fields in the original data.
Considering the shortcomings of classifying industries in a standard way or obtaining the operating countries for multi-national or lesser known organizations not present in DBpedia, we add the annotated fields from the dataset in the knowledge graph as the industries and countries for the entities in addition to the ones obtained from DBpedia.
In order to maintain the integrity of the dataset, we omit the rows where the victim is not specifically reported (i.e. various victims in multiple countries) or the countries or industries are not exact in a similar manner.
The resulting nodes are resolved by comparing them to one another by using string similarity and sufficiently similar nodes are joined.
Processing the dataset using these methods resulted in a knowledge graph of 12,966 nodes and 18,476 edges.
The filtered and processed data leaves us with 6,825 attack SVO triples.

We construct the risk measure by first binning and ordering the recorded attacks and the triples into daily bins, after which we calculate the number of attacks towards the entities connected to the central nodes.
The risk levels for each central node were calculated using the formula in Eq.~(\ref{Eq:centralrisk}).
For each non-central entity illustrated in the resulting knowledge graph, we construct the averages of the first and second neighbor countries and industries in the graph into sets of four variables for each day, i.e. $\{\overline{r_e(C)}, \overline{r_e(I)}, \overline{r_e(c)}, \overline{r_e(i)}\}$.
We split the dataset into ``attack days'' and ``non-attack days'' by letting the values for the attack days to be the values for the previous time step and sampling a non-attack day as a random day between the beginning of the dataset and the recorded attack date. 
By this process we obtain a dataset with 11,028 observations, with equal amount of points in both classes.

%logistic regression and PCA
To further interpret the usefulness of the constructed risk variables we perform a logistic regression and dimensionality reduction on the dataset of attack days and non-attack days.
The objective of this analysis is to investigate the separability of the two classes and the relationship between the risk variables.
These methods were implemented with the Scikit-learn library \cite{scikit-learn}.
For the logistic regression classifier we split the constructed dataset into a training set and testing set with a 60 to 40 ratio at random.
Fitting the logistic function to the training data allows us to evaluate the feasibility of the risk measures, even though the objective of this framework is not to predict the exact attack dates.
By fitting the classifier we also obtain the weights for the variables which we can interpret in terms of their importance and mutual relationship.
The last part of our evaluation was perfomed using a principle component analysis (PCA) on the dataset to portray the dataset in two dimensions.

\section{Results and analysis on an experimental dataset}

\begin{figure}[h]
    \centering
    \includegraphics[width=0.75\textwidth]{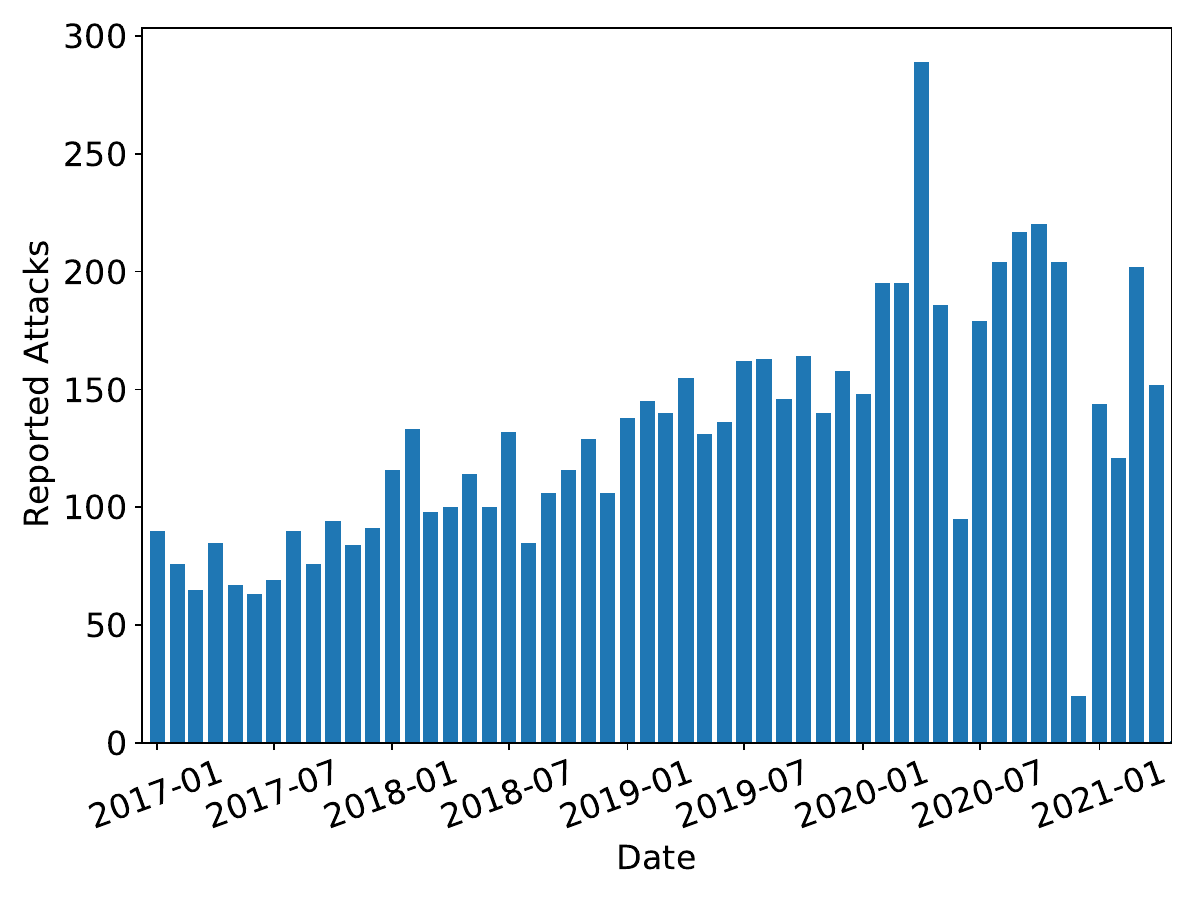}
    \caption{Number of monthly incidents in the filtered dataset. Records that did 
    not produce a single coherent SVO (subject-verb-object) triple for the attack were omitted to produce a coherent knowledge graph. The time of occurrence of an event (a triple) can be wrongly recorded or reported long after the attack. It is notable from the number of attacks that the reporting is not uniform and some months are much less populated than others, resulting from human error or bias.}
    \label{fig:timeseries}
\end{figure}

\begin{figure}[h!]
    \centering
    \includegraphics[width=0.45\textwidth, trim=50 440 200 0, clip]{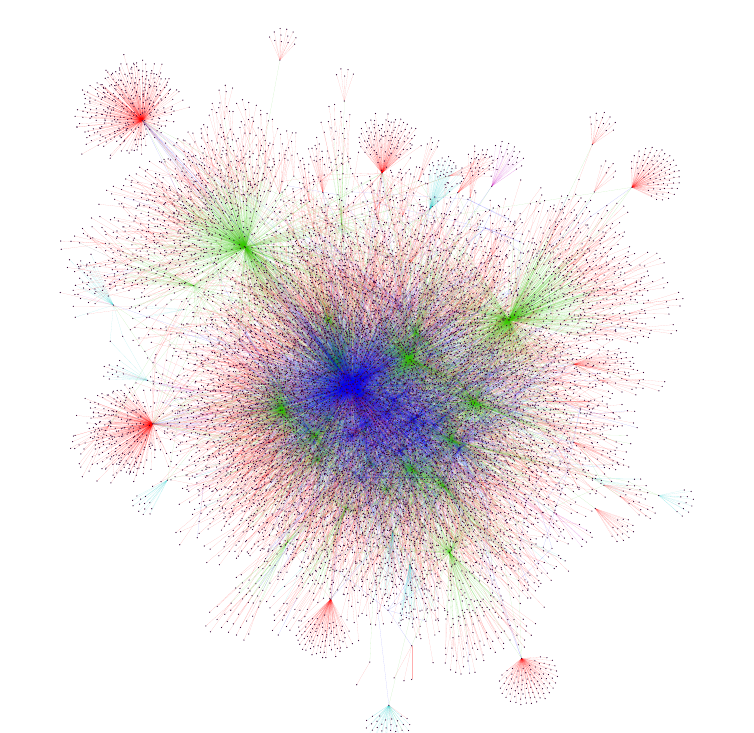}
    \includegraphics[width=0.45\textwidth, trim=100 500 200 0, clip]{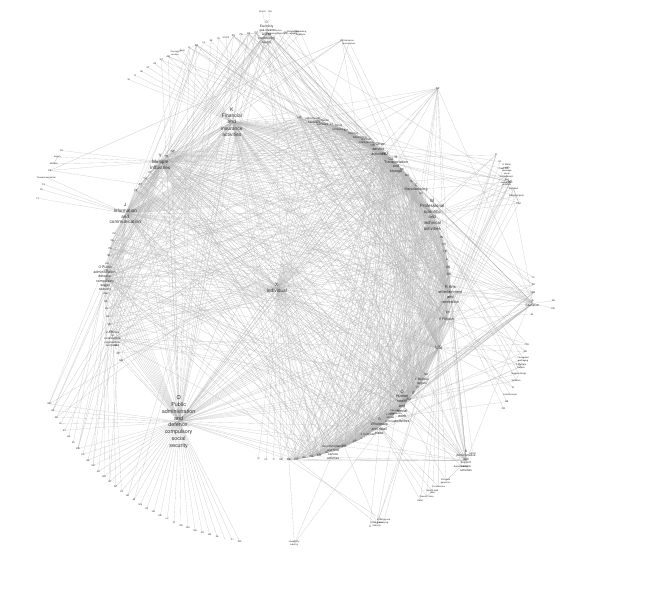}
    \caption{(Left) The resulting knowledge graph from the Hackmageddon dataset
    of 2017-2020. The edges are coloured according to the interaction in the related triples such that red edges represent the attack triples (attackedBy), blue edges represent hasCountry, green edges represent hasIndustry, purple edges represent hasProduct triples, and turquoise edges represent hasParent triples. (Right) The projection of the central nodes used in the construction of the entity risk measures. The projection is constructed by linking central nodes sharing common neighbors such that the weight of every link is uniform regardless of the number of common neighbors.}
    \label{fig:hacknet}
\end{figure}

\begin{figure}[h!]
    \centering
    \includegraphics[width=0.8\textwidth, trim=100 500 200 0, clip]{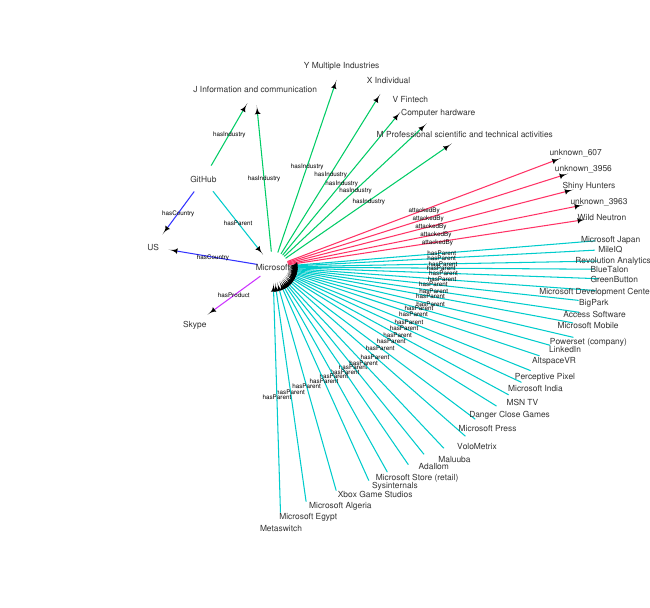}
    \caption{An example subset of the knowledge graph showing an egocentric network from the company Microsoft. The central entity is connected to the reported malicious entities (red links), the industries reported in the dataset as well as the ones obtained from DBpedia (green links), country (blue link), products (purple links) and child companies (turquoise links). The network is a subset of the knowledge graph with nodes and links of a single step from the focal node. 
    The central nodes of this subset are connected to the focal node by green and blue links.}
    \label{fig:case}
\end{figure}

The frequency of reported attacks in the combined dataset is depicted in Fig. \ref{fig:timeseries}. 
The number of reported attacks per month shows an increasing trend but teh months of June and December in 2020 have significantly lower number of incidences, hinting that the reporting in the data can be incomplete. 
The nodes with the highest degree are industries (public sector, healthcare) and countries (US, UK). 
The non-central nodes with the highest degree are the tech giants such as Google and Amazon and the users of their products such as the Android operating system.
As we use the industry fields from the annotated dataset in addition to the ones obtained using DBpedia Spotlight, the network consists of a single connected component.
The finalized knowledge graph based on the dataset from Hackmageddon cyberattack timelines from January 2017 to April 2021 and extraneous information from DBpedia is shown in Fig. \ref{fig:hacknet} and a more descriptive subset of the same graph is shown in Fig. \ref{fig:case}.

\begin{figure}[ht!]
    \centering
    \includegraphics[width=0.95\textwidth]{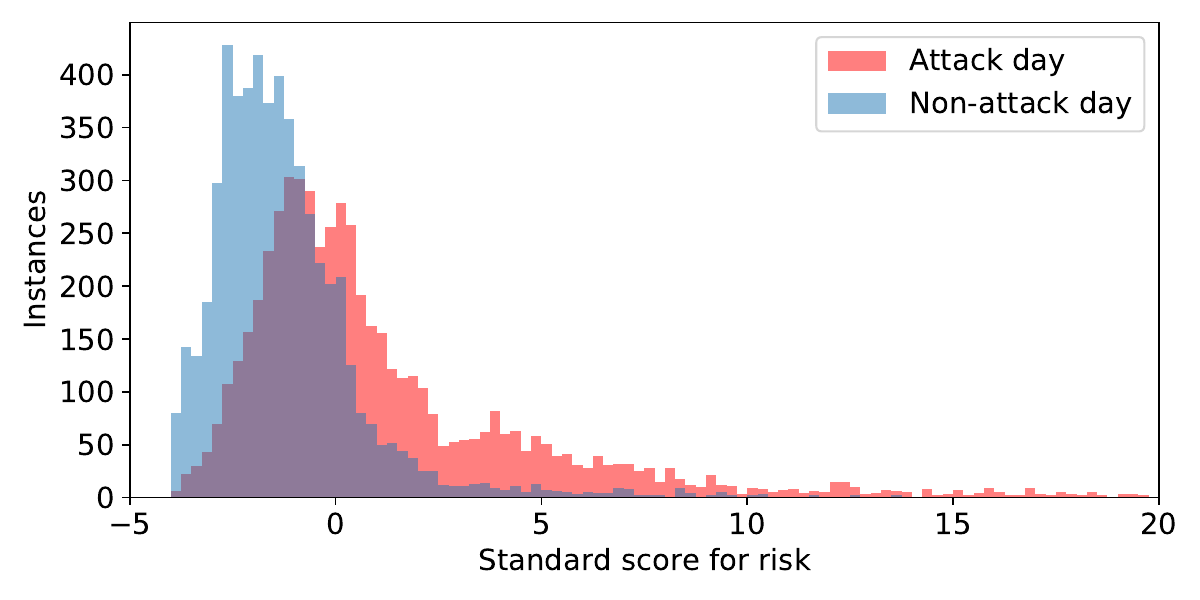}
    \includegraphics[width=0.45\textwidth]{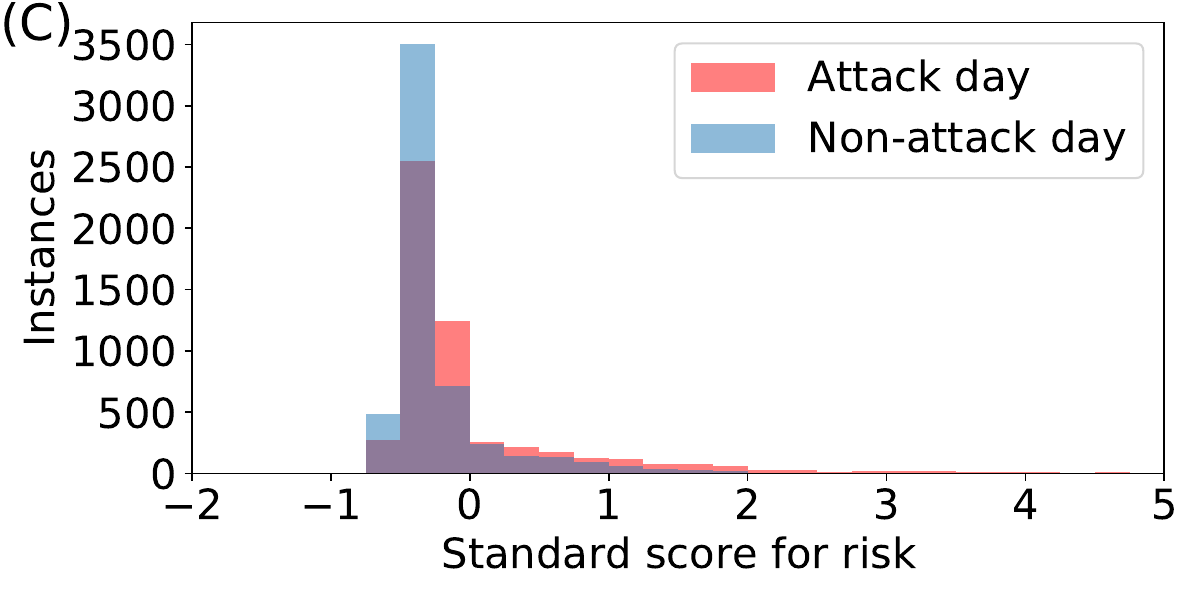}
    \includegraphics[width=0.45\textwidth]{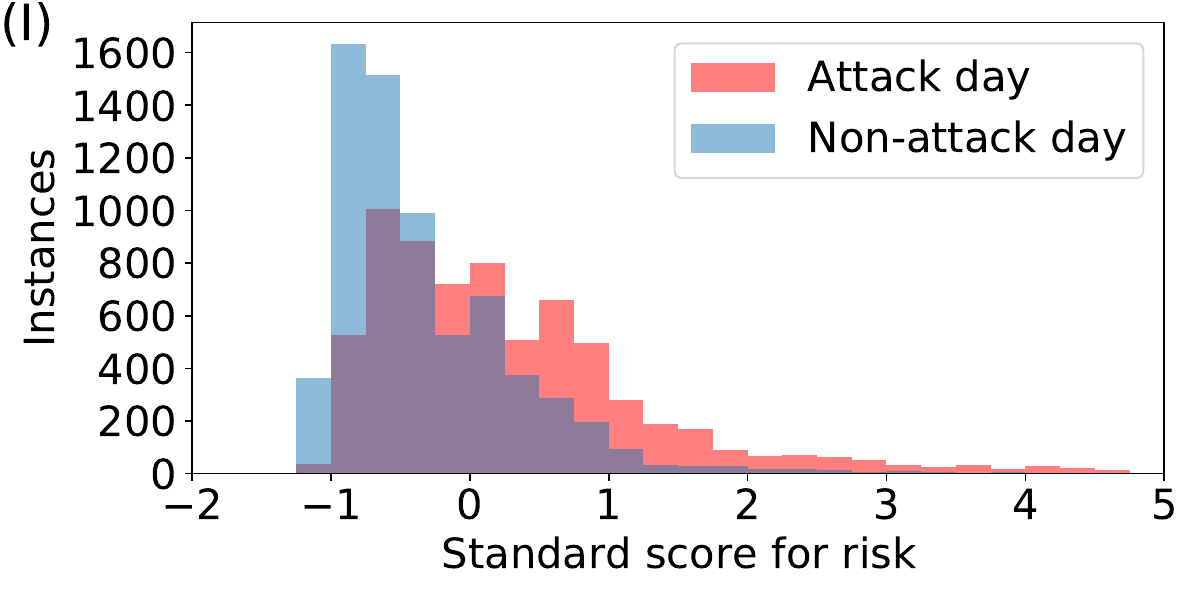}
    \includegraphics[width=0.45\textwidth]{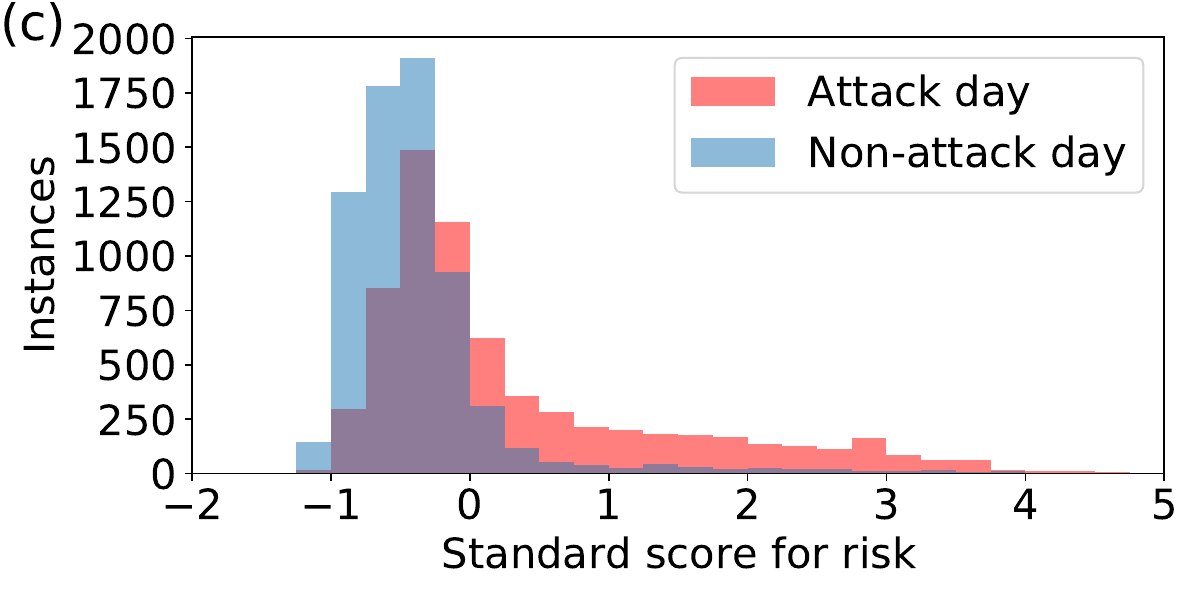}
    \includegraphics[width=0.45\textwidth]{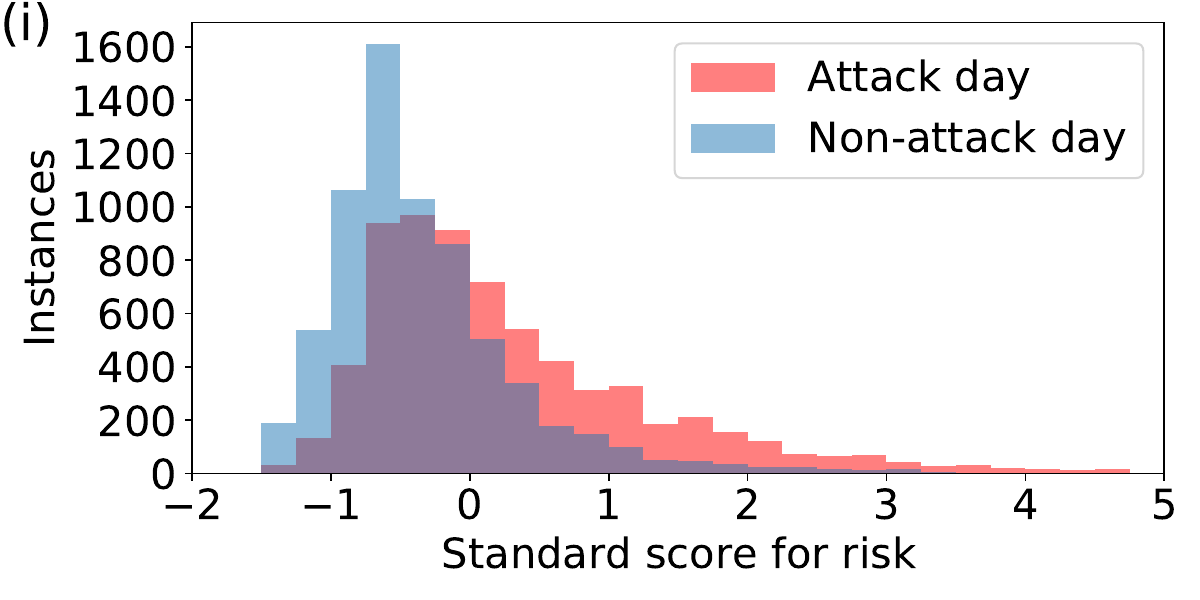}    \caption{The distributions of resulting risk levels of attack and non-attack days in the knowledge graph. (Top) The sum of the four variables, (Second row) The standard score for risk in the first neighbor country node $C$ and the standard score for the average risk in the first neighbor industry nodes $I$. (Third row) The standard score for average risk in the second neighbor countries $c$ and second neighbor industries $i$. The values are standard scores for recorded attack days and equal number of sampled non-attack days to the same entity before the attack day. }
    \label{fig:risklevels}
\end{figure}

The standardized distributions for the four different variables to the right in Eq.~(\ref{Eq:risk-variables}) are shown in Fig. \ref{fig:risklevels}.
Overall, the distributions between the classes seem to differ from one another, the attack distribution having a longer tail and more positive mean.
It is notable that the distribution of the first neighbor's (country) risk is very similar between the attack days, whereas the second neighbor's risk shows a difference between the attack days and non-attack days.
The differences between attack days and sampled attack days in the distributions or risk values show that there is some commonality within the classes.

\begin{figure}
    \centering
    \includegraphics[width=0.40\textwidth]{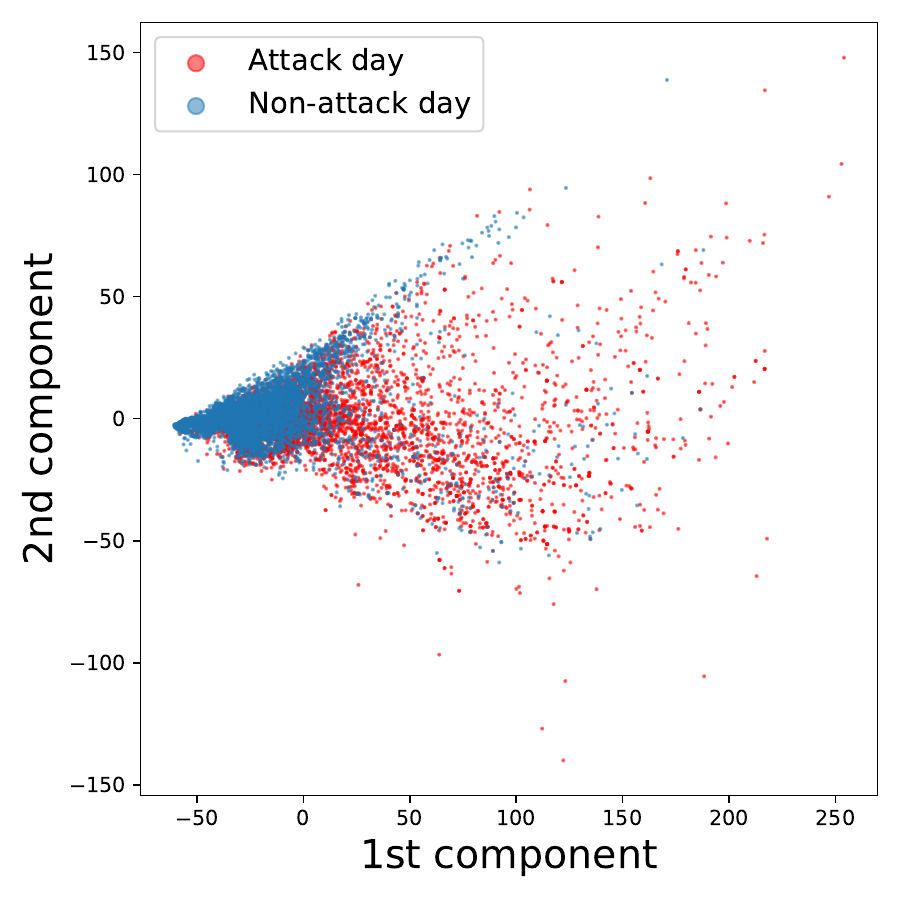}
    \includegraphics[width=0.55\textwidth]{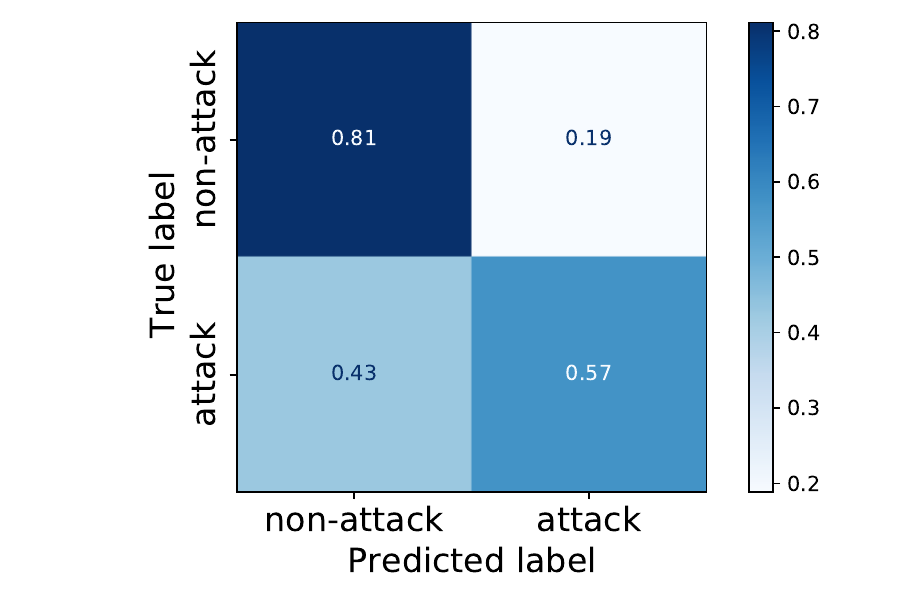}
    \caption{
    (Left) First two components of PCA dimensionality reduction on the risk data consisting of the four variables (average risk values of the first and second countries and industries). 
    The first component resulted in positive weights for the secondary central nodes and the second component resulted in negative weights for the same variables. 
    The difference shown in the plot hints that the method produces higher risk levels for the dates when the attacks occur. 
    (Right) Confusion matrix obtained by testing a logistic regression classifier on the proposed risk measures. The data was split into training and testing sets with a 60 to 40 ratio. The overall accuracy of the classifier was 0.69 and F1 score was 0.65.
    }
    \label{fig:confmat}
\end{figure}

Training a logistic regression classifier with a standardized training and validation set constructed from the data results in a 69\% accuracy, which shows that there is, indeed, some relationship between the attacks in the network, at least in a temporal sense appearing as burstiness.
The coefficients for the logistic regression (see Table. \ref{table:coeffs}) show that the first neighbour country has a very minor weight in the classifier function. 
The corresponding distribution in Fig. \ref{fig:risklevels} reinforces this as the two classes are highly overlapping.
The interesting fact is that the coefficient of second neighbor country is the highest, which could be interpreted as some countries being the catalysts for chains of attacks.
The confusion matrix for the logistic regression is depicted in Fig. \ref{fig:confmat}, showing the fractions of correctly and incorrectly predicted labels, 1 being the label ``attack day'' and 0 being the label ``non-attack day''.
As one would expect, the accuracy for correctly predicting ``non-attack days'' is higher than correctly predicting the ``attack days'' due to the differences in the distributions of the constructed variables.

Performing a dimensionality reduction in the form of principle component analysis (See Fig. \ref{fig:confmat} left panel) results in components explaining 94\% of the variance (83\% and 11\% for the two components).
The component weights are shown in Table \ref{table:coeffs}.
These weights can be interpret as two different risk factors, industry-based and system-based risk.
The industry-based risk in this situation can be reasoned from the higher factors for the industry nodes in the first principal component and the system-based risk can be considered due to negative factors to all but second neighbour industry risk $i$.
Also judging from the Figure \ref{fig:confmat}, the second component differentiates between risk from the secondary industries and the first neighbor nodes as well as the second neighbor country.
The small factors for first neighbor country $C$ are apparent from the overlap in the variable's distribution shown in Figure \ref{fig:risklevels}.

\begin{table}[h]
\caption{PCA component weights and coefficients from fitting a logistic regression to the data. The variables are notated as first neighbor country ($C$), first neighbor industry ($I$), second neighbor country ($c$) and second neighbor industry ($i$).
\label{table:coeffs}}
\begin{center}
 \resizebox{0.9\textwidth}{!}{
\renewcommand{\arraystretch}{1.5}
\begin{tabular}{|c|c|c|c|}
\hline
\textbf{Variable}  & \textbf{PCA 1st component}  & \textbf{PCA 2nd component}  & \textbf{Logistic regression coefficient}  \\ \hline
\hline
$C$ & 0.075 & -0.127 & -0.004\\ \hline
$I$ & 0.516 & -0.737 & 0.025\\ \hline
$c$ & 0.176 & -0.377 & 0.039\\ \hline
$i$ & 0.835 & 0.547 & 0.007\\ \hline
\end{tabular}
}
\end{center}
\end{table}
%[[-0.00436351  0.02494743  0.03851992  0.00690015]] [0.19681644]
%pca
%[[ 0.07501786  0.51634253  0.17580269  0.83477909]
%[-0.12741183 -0.7369624  -0.37665411  0.54661167]]

\section{Discussion}

As a whole, the advantages of the presented framework are its generality, explainability and expandability.
The generality allows us to consider a large variety of different types of attacks and entities with limited information and resources.
The information used to construct the strategic level graph with high abstraction is mostly available in online sources such as DBpedia and the extraction of the victim-attacker triple from unstructured data can be performed efficiently using the methods presented in this paper.
The ontology is human-readable and easy to explain and can thus serve as a tool for communicating a large number of events and investigations with other people, such as the management responsible for organizational decision making. As the framework can be used to semi-automatically produce a contextual situational picture of the cyberspace and compute levels of risk for entities and industries, it could also serve as a tool for companies with limited resources on cyber intelligence and analysis.
The framework can be extended to include information on software vulnerabilities, software used by the entities, and importance of entities in various supply-chain systems, moving the knowledge graph towards a more operational scope.
However, the availability of such information is restricted and for the scope of this study we decided to keep the network structure human-readable by having only the essential nodes for describing general entities and incidents and relationships between them.

The analysis of the risk measure has shown that there can be some level of temporal and structural correlation between the different recorded attacks.
The distributions between the ``attack'' and ``non-attack days'' in the dataset differ from each other to a degree (See Fig.~\ref{fig:risklevels}) and performing a logistic regression classification on the produced knowledge graph dataset yields a decent accuracy (See Fig. \ref{fig:confmat}).
This reinforces the usefulness of our strategic level ontology, which assumes that similar entities have some common factors that are not always publicly reported and that similar companies are often targeted during some period of time.
The relationship between different entities in the knowledge graph can be more complex than just surface level similarities and contain hidden variables, such as the used systems and protocols, which could explain some of the pathways between the various entities that are connected to different central nodes.
Correlations between attacks and attacked entities in the knowledge graph can also be because the attackers focus on certain type of entities for their own reasons.
The risk measures presented here are intended for evaluating our framework rather than investigating the real life risk.
The results show that our framework has potential in formulating a measure of risk in addition to the capabilities on visualizing a large dataset for situational awareness and investigation.

There are also limitations to this framework.
The risk measure used in this study is based on the network structure of the resulting knowledge graph and thus the design choices have high impact on the variables constructed.
%The classification and prediction of cyberattacks with information limited to open source reports, is likely to result in an uncertainty. 
The results can be biased and affected by numerous sources, such as human error and bias in the reporting of the incidents and collecting the incidents to the dataset.
%beginning of edit
The constructed knowledge graph is hardly a ground truth of all cyberattacks due to lack of reporting or detection. 
The language of the collected dataset can impose a limitation on geographical areas where the reporting language is different from the one used in the implementation of the framework.
%On the other hand, in a practical scenario the analyst using such framework is choosing the source material to their interest and available language capabilities.
%end of edit
Other biases can rise from the accuracy of processing the unstructured data into a knowledge graph as well as the types of extra information added to the graph, such as the types of industry nodes.
The generality of the industries have a direct effect on the structure of the graph and the related properties.
Also, the accuracy for recording and reporting a cyber-related incident can vary in public online sources. The discovery of an attack can be late and thus the accurate time might not be reported correctly. This is a direct limitation to the estimation of risk in the framework. For instance, the dates reported in the empirical dataset used in this study vary in accuracy with some dates reported days later than the incident in question.

Within the information extraction pipeline there are two primary sources of possible inaccuracies. The first involves information from the dependency parser. This was done by using spaCy’s trained pipelines, and therefore, indirectly depended on the performance of the in-built neural network predictions. The second source of inaccuracy is the scoring method used for ranking of triples. While we achieve a moderate accuracy of 60$\%$ (and 83$\%$ for target to lie within the top three ranks) by this method, the results indicate the importance of inclusion of the target score. Just using the order of appearance or frequency lowers the accuracy by 10$\%$. Also, as noted earlier the baseline accuracy would be much smaller. 

%Evaluation and discussion of risk

% discussion

% discussion

%\st{The design and dataset in this study were chosen due to the ease of performing preliminary investigations with the framework and the limitations of open and annotated data sources. In addition to the human errors and design choices having a language specific tools for NLP, causes the information retrieval to be restricted to a certain part of the world, which in itself limits the amount of information available and the types of entities and reported incidents. On the other hand, in an application of the framework the analyst is eventually choosing the source material of their interest with the knowledge of the scope of their investigation.}

\section{Conclusions}

In this study we have presented a novel knowledge graph based framework for constructing a strategic level mapping of the current and past cyberattacks from unstructured reports in the open online sources and demonstrated the capabilities of the resulting knowledge graph in terms of communicating events and constructing measures for risk.
The aim of this framework is to structure textual data into computer-readable and computable form, facilitate measures for risk and help expert analysts to process and view a large amount of reports in an automated manner.
The pipeline combines methods and techniques from NLP and complex networks, starting with scraping and retrieving of articles from online sources, extracting relevant entities and the correct subject-verb-object or SVO-triples on the attacked entities and the attacking actors, and finalizing by constructing a knowledge graph with an ontology consisting of five types of nodes and relationships (see Fig.~\ref{fig:onto}).
We have implemented the pipeline and the related algorithms in Python 3.7 programming language and created a knowledge graph using the pre-annotated dataset from Hackmageddon that contains over 7000 recorded attacks between January 2017 and April 2021 (See Fig.~\ref{fig:hacknet}).
With this knowledge graph we have also  constructed a measure of risk, which is based on a decaying time-based function and the network structure of the knowledge graph.

%To summarize, we have proposed a novel framework for structuring records on cyber attacks and demonstrated the capabilities of the resulting knowledge graph in terms of communicating events and constructing measures for risk.
We believe that the methods and results of this study can help cyber-analysts to perform their investigations more efficiently in the future as the amount of new information is increasing faster than the number of experts available at any time.
In our future research, we plan to improve the methods for information extraction from unstructured sources for better accuracy and generalization, which would improve the reliability and validity of the knowledge graph as well as provide a possibility for better automation in terms of facilitating the framework as a continuous process.
Constructing language-agnostic tools for this task would also solve the problem of having a limited focus on certain parts of the world.
As discussed previously, adding new information from other sources, such as system information of entities and various vulnerability databases, could increase the accuracy of the risk model, should such information be available. This would also allow to conduct simulations and ``what-if'' type scenarios on the knowledge graph, possibly being able to show more microscopic trends or campaigns. In addition to high-level scenarios, such as common infrastructure, one could utilize the vast work done in the field of models and simulations for vulnerability analysis (e.g. \cite{ficco2017simulation, kavak2021simulation}). Also, joining the system level technical information and accurate industry information to the framework could allow categorization of the events into different aspects of the society such as political, economical and military-operations.
\bibliographystyle{unsrtnat}
\bibliography{refs}

\section*{Author Contributions}
TT and KB constructed the strategic ontology and the knowledge graph model.
KB constructed the NLP method for extracting the SVO-triples from the textual data.
TT collected and constructed the knowledge graph and the related risk measures and created the figures for the manuscript.
TT and KB wrote the manuscript with the support and feedback of KK, ML, PJ and AC.

\section*{Acknowledgments}
TT, KB, ML and KK acknowledge research project funding from Cyberwatch Finland. TT acknowledges funding from the Vilho, Yrjö and Kalle Väisälä Foundation of the Finnish Academy of Science and Letters.

\end{document}